\documentclass[twocolumn,showpacs,floatfix,prc]{revtex4}
\usepackage{dcolumn}
\usepackage{bm}
\usepackage{color}

\usepackage{graphicx}
\usepackage{amsmath}
\usepackage{amssymb}

\def\gsim{\buildrel > \over {_{\sim}}}

\def\beq{\begin{equation}}
\def\eeq{\end{equation}}
\def\be{\begin{eqnarray}}
\def\ee{\end{eqnarray}}

\begin{document}
\title{Two particle-two hole final states in quasi elastic neutrino-nucleus interactions}
\author{Omar Benhar$^{1}$}
\altaffiliation[on leave from ]{INFN and Department of Physics, ``Sapienza'' Universit\`a di Roma, I-00185 Roma, Italy.}
\author{Alesssandro Lovato$^{2,3}$}
\author{Noemi Rocco$^{4,5}$}

\affiliation
{
$^1$ Center for Neutrino Physics, Virginia Polytechnic Institute and State University,
Blacksburg, VA 24061, USA  \\
$^2$ Argonne Leadership Computing Facility, Argonne National Laboratory, Argonne, IL 60439, USA\\
$^3$ Physics Division, Argonne National Laboratory, Argonne, IL-60439, USA \\
$^4$ INFN, Sezione di Roma, I-00185 Roma, Italy \\ 
$^5$ Department of Physics, ``Sapienza'' Universit\`a di Roma, I-00185 Roma, Italy
}
\date{\today}
\begin{abstract}
The excitation of two particle-two hole final states in neutrino-nucleus scattering, not taken into account in Monte Carlo simulations for data analysis,  has been 
advocated by many authors as the source of the excess cross section observed by the MiniBooNE collaboration in the quasi elastic 
channel. We analyse the mechanisms leading to the appearance of  two particle-two hole states, and show that
interference between the amplitudes involving one- and two-nucleon currents, not consistently included in existing calculations, 
plays a major role.  A novel approach allowing to treat one- and two-nucleon current contributions on the same footing is outlined.

\end{abstract}
\maketitle

Experimental studies of neutrino-nucleus interactions  
 carried out over the past decade \cite{K2K,MiniBooNE_Q2dist,nomad,MiniBooNE_d2sigma} have
 provided vast evidence of the inadequacy of the Relativistic Fermi Gas Model (RFGM), routinely 
 employed in event generators for data analysis, to account for both the complexity of nuclear dynamics and the variety 
of reaction mechanisms -- other than single nucleon knock out -- contributing to the observed cross section.


A striking manifestation of the above problem is the large discrepancy between the predictions of Monte
 Carlo simulations and the double differential charged current quasi elastic (CCQE) cross section 
 measured by the MiniBooNE collaboration using a carbon target  \cite{MiniBooNE_d2sigma}. 

As pointed out by the authors of Ref. \cite{coletti}, improving the treatment of nuclear effects, 
which is now acknowledged as one of the main sources of systematic uncertainty \cite{T2K}, 
will require the development of a {\em comprehensive} and {\em consistent} description of neutrino-nucleus 
 interactions, validated through comparison to the large body of accurate electron-nucleus 
 scattering data \cite{PRD,RMP}.
  
The main difficulty involved in the generalisation of the approaches successfully employed to analyse 
electron scattering 
to the case of neutrino interactions stems from the fact that, while in electron scattering 
the beam energy is fixed, in neutrino scattering the measured cross section is obtained by averaging over
different beam energies, distributed according to the neutrino flux. As a consequence, a measurement of the energy of the
outgoing charged lepton {\em does not} specify the energy transfer to the nuclear target, which 
determines the dominant reaction mechanism.  As shown in Ref. \cite{benhar:nufact}, the 
MiniBooNE double differential cross section corresponding to a specific muon energy bin turns out to receive comparable 
contributions from different mechanisms, which must be all taken into account.

Many authors have suggested that the excess CCQE cross section observed by the MiniBooNE collaboration is to be ascribed
to the occurrence of events with two particle-two hole (2p2h) final states, not included in Monte Carlo simulations  \cite{martini, coletti,nieves}. 
A consistent description of these processes
within a realistic model of nuclear dynamics requires that all mechanisms leading to their appearance~--~Initial State Correlations (ISC) among 
nucleons in the target ground state,
Final State Correlations (FSC) between the struck nucleon and the spectator particles, and interactions involving two-nucleon meson-exchange 
currents (MEC)~--~be included. 
In existing calculations carried out in the kinematical region
relevant to MiniBooNE analysis  \cite{martini,nieves}, however,  the initial and final nuclear states are described within the 
Independent Particle Model (IPM),  the deficiencies of which have been most clearly highlighted over fifty years ago by Blatt and Weisskopf, in their 
classic Nuclear Physics book  \cite{BW}.
  
In this Letter, we analyse the mechanisms leading to the appearance of 2p2h final states, and argue that interference between the amplitudes 
involving one- and two-nucleon currents, not consistently accounted for in  Refs.~\cite{martini,nieves}, may play a critical role. 
We also outline a novel approach, based on a generalisation of the factorisation {\em ansatz} implied in the impulse approximation (IA) scheme, 
 allowing one to treat one- and two-nucleon current contributions on the same footing.

The nuclear electroweak current, determining the nuclear response to electron and neutrino interactions, 
can be written as a sum of one- and two-nucleon contributions according to (see, e.g., Ref. \cite{2NC})
\beq
\label{nuclear:current}
J_A^\mu= \sum_i j^\mu_i+\sum_{j>i} j^\mu_{ij} \ .
\eeq
The one-body operator $j_i^\mu$ describes interactions involving a single nucleon, and can be expressed in terms of the vector and 
axial-vector form factors. The two-body current $j^\mu_{ij}$, on the other hand, accounts for processes 
in which the beam particle couples to the currents arising from meson exchange between two interacting nucleons.

It is very important to realise that, in scattering processes involving {\em interacting} many-body systems, 2p2h final states 
can be produced through the action of {\em both} one- and two-nucleon currents. 
Within the IPM, however, in which interaction effects are described in terms of a mean field, 2p2h states can {\em only} be excited by two-body 
operators, such as those describing MEC.
In order for the the matrix element of a one-body operator between the target ground state and a 2p2h final state to be non vanishing, 
the effects of dynamical nucleon-nucleon (NN) correlations, ignored altogether in the IPM picture, must be included in the description of 
the nuclear wave functions. 

Correlations give rise to virtual scattering between target nucleons, leading to the excitation of the participating 
particles to continuum states. The ISC contribution to the 2p2h amplitude arises from processes in which the 
beam particle couples to {\em one} of these high-momentum nucleon.  The FSC contribution, on the other hand, originates from scattering 
processes involving the struck nucleon and one of the spectator particles, that also result in the appearance of  2p2h final states.  

In the kinematical region corresponding to moderate momentum transfer, typically $|{\bf q}| < 400 \ {\rm MeV}$, in which non relativistic approximations 
are expected to be applicable,  ISC, FSC and MEC can be consistently described within advanced many-body approaches based on realistic models of nuclear dynamics,  
strongly constrained by the properties of the {\em exactly solvable} two- and three-nucleon systems  \cite{2NC}.
The results of  non relativistic calculations, while not being directly comparable to experimental data at large momentum 
transfer, can provide valuable insight on the interplay of the different mechanisms leading to the excitation of 2p2h final states.

The authors of Ref. \cite{lovato12C} have recently reported the results of an accurate calculation of the sum rules of the electromagnetic 
response of carbon in the longitudinal and transverse channels, carried out within the Green's Function Monte Carlo (GFMC) computational scheme. 
Exploiting the completeness of the set of final states entering the definition of the nuclear inclusive  cross section, these sum rules can be easily related to the
energy-loss integrals of the longitudinal and transverse components of the tensor describing the target response to electromagnetic 
interactions \cite{RMP}.  

Choosing the $z$-axis along the direction of the momentum transfer, ${\bf q}$, the transverse sum rule can be written in the form
\beq
\label{sr1}
S_T({\bf q}) = \int d \omega S_T({\bf q},\omega)  \ ,
\eeq
where 
\beq
\label{sr2}
S_T({\bf q},\omega) = S^{xx}({\bf q},\omega) + S^{yy}({\bf q},\omega) \ ,
\eeq
with ($\alpha, \ \beta =$ 1, 2, and 3 label the $x$- $y$- and $z$-component of the current, respectively)  
\beq
\label{sr3}
S^{\alpha \beta}=\sum_N \langle  0| J_A^\alpha | N \rangle  \langle  N | J_A^\beta |   0 \rangle \delta(E_0+\omega-E_N) \ .
\eeq
In the above equation, $| 0 \rangle$ and $| N \rangle$ denote the initial and final nuclear states, the energies of which are $E_0$ and 
$E_N$. The generalisation of Eqs. \eqref{sr1}-\eqref{sr3} to the case of charged current weak interactions is discussed in Ref. \cite{LovatoNPA}. 

We have employed the approach of Ref. \cite{lovato12C} to pin down the contribution of the terms arising from interference between 
correlations 
and MEC to the transverse sum rule, which is long known to  be strongly affected by processes involving two-nucleon currents.

\begin{figure}[h!]
 \includegraphics[scale= 0.5]{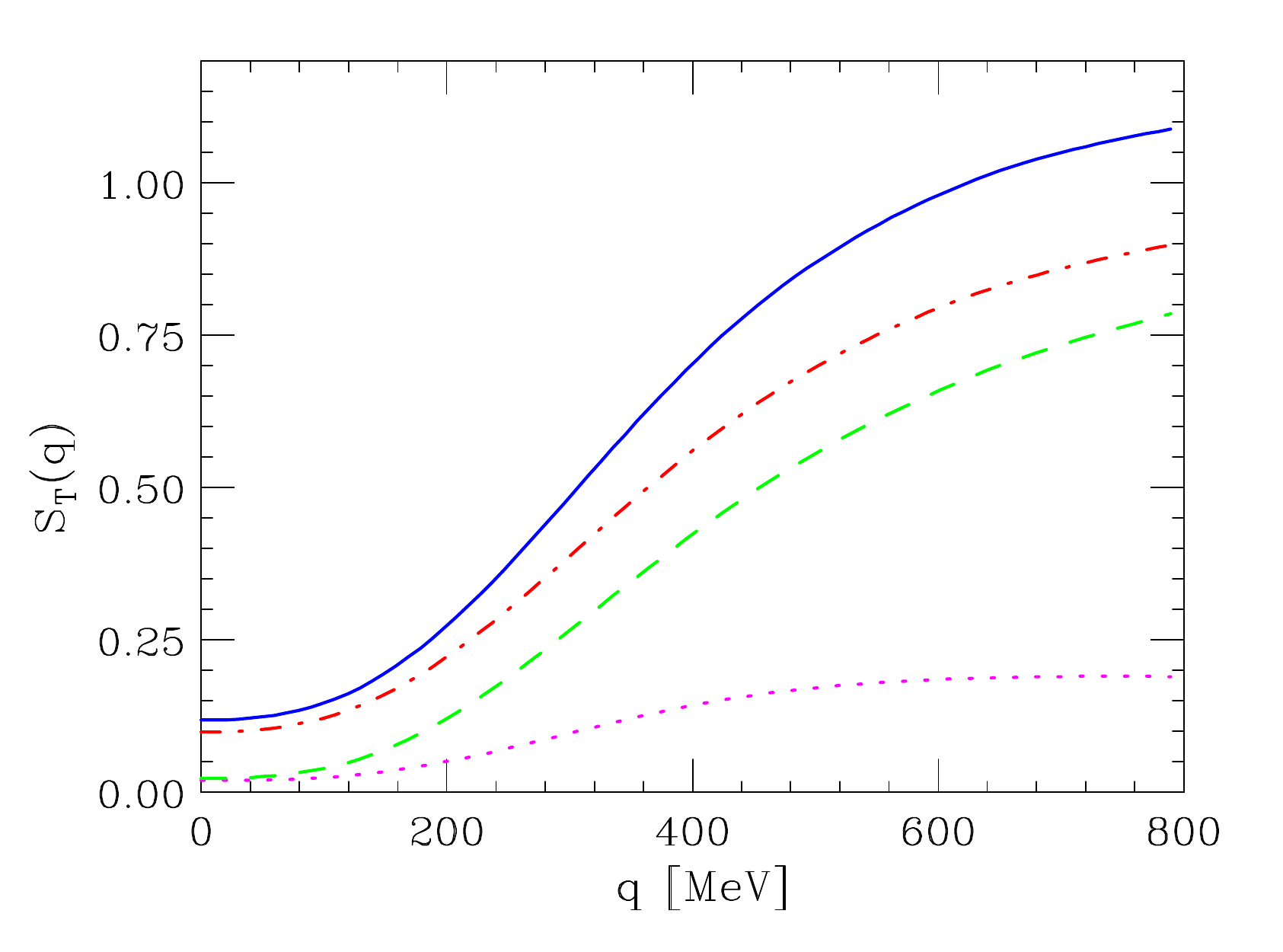} 
\vspace*{-.2in}
\caption{Sum rule of the electromagnetic response of carbon in the transverse channel. The dashed line 
shows the results obtained including the one-nucleon current only, while the solid line corresponds to the 
 full calculation. The dot-dash line represents the sum rule computed neglecting interference 
terms, the contribution of which is displayed by the dotted line. The results are normalised so that the dashed line approaches 
unity as $|{\bf q}| \to \infty$. Monte Carlo errors bars are not visible on the scale of the figure.}
\label{sumrule}
\end{figure}

The results of numerical calculations, displayed in Fig.~\ref{sumrule}, clearly show that interference terms provide a sizeable fraction of the sum rule.
At momentum transfer $|{\bf q}| \gsim 300$ MeV, their contribution turns out to be comparable to -- in fact even larger than -- that obtained squaring 
the matrix element of the two-nucleon current.

Within the approach of Refs. \cite{martini,nieves}, based on the IPM description of the nuclear initial and final states, 
interference terms are generated by adding {\em ad hoc} contributions to the two-body current \cite{torino}. However, this 
procedure does not properly account for correlations arising from the strong repulsive core of the NN interaction.
Furthermore, it disregards correlations among the spectator particles altogether.   

\begin{figure}[h!]
\includegraphics[scale=0.45]{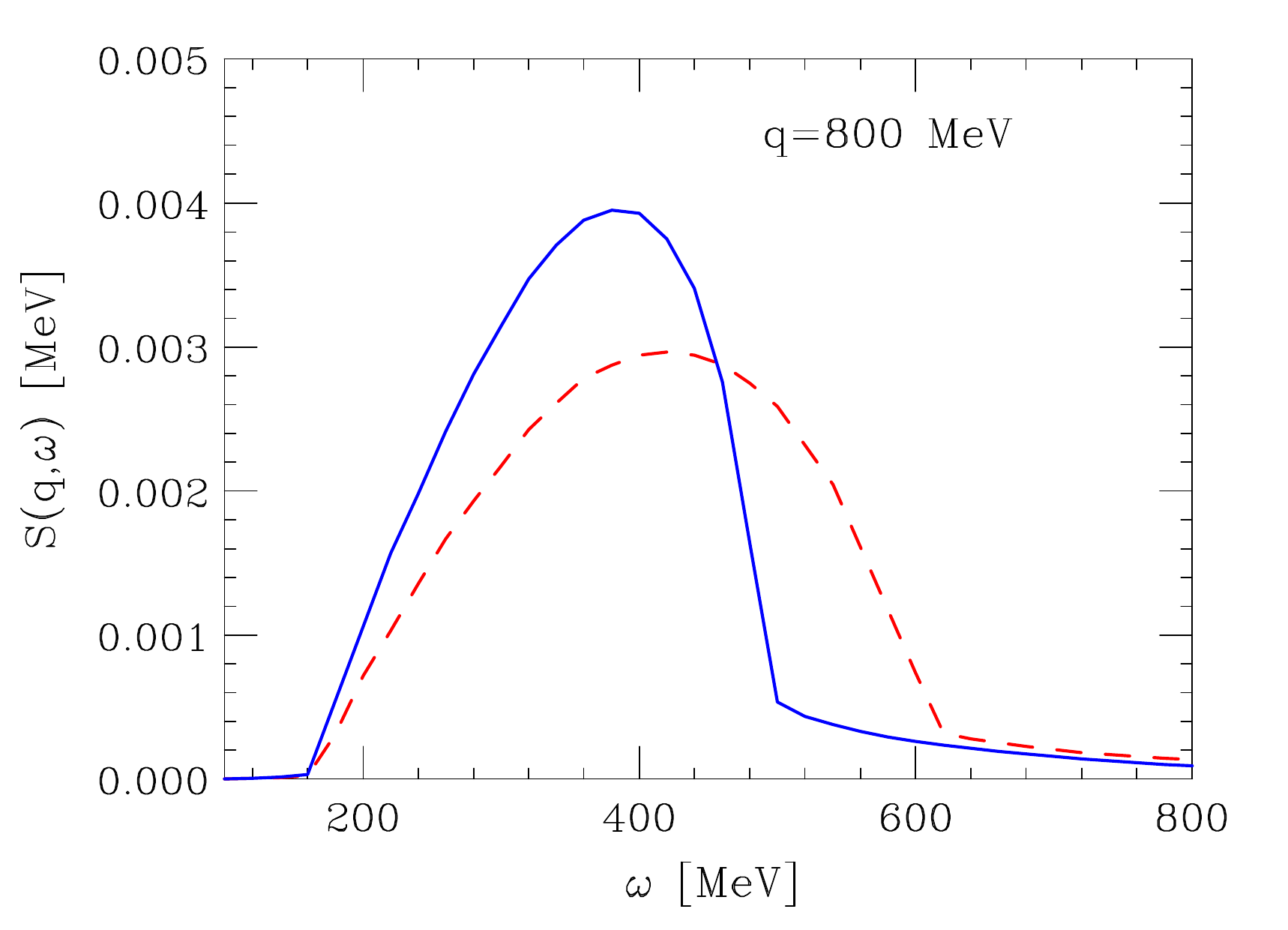}
\vspace*{-.1in}
\caption{Nuclear matter response to a scalar probe delivering momentum $|{\bf q}| = 800$ MeV. The solid and dashed lines have been obtained using 
relativistic and non relativistic kinematics, respectively.}
\label{relkin}
\end{figure}

The results of Fig.~\ref{sumrule} clearly point to the need for a consistent treatment of correlations and MEC within a 
formalism suitable for application in the kinematical regime in which non relativistic approximations are known to fail. The relativity
issue is of paramount relevance to the analysis of neutrino data, because the mean momentum transfer of CCQE events 
obtained by averaging over the MiniBooNE \cite{MiniBooNE_d2sigma} and Miner$\nu$a \cite{Minerva} neutrino fluxes
turn out to be $\sim 640$ and $\sim 880$ MeV, respectively. Comparison between the solid and dashed lines of Fig.~\ref{relkin}, showing the nuclear matter response 
to a scalar probe delivering momentum $|{\bf q}| = 800$ MeV,  demonstrates that 
the non relativistic approximation fails to predict both position and with of the quasi elastic bump. The calculations have been carried out using the 
formalism described in Ref. \cite{RMP}.

The effects of ISC on the nuclear cross section at large momentum transfer can be taken into account within the formalism 
based on the IA using {\em realistic} spectral functions \cite{PKE,LDA}. The IA scheme rests on the assumptions that at momentum transfer such that
$|{\bf q}|^{-1} \ll d$, $d$ being the average separation distance between nucleons in the target nucleus, the contribution of the two-nucleon current can be 
disregarded and the final state  $| N \rangle$ of Eq.~\eqref{sr3} can be written in the factorized form
\beq
\label{fact1}
| N \rangle = | {\bf p}\rangle \otimes | n_{A-1}, {\bf p}_n \rangle \ ,
\eeq
where the state $|{\bf p}\rangle$ describes a non interacting nucleon carrying momentum ${\bf p}$, while  $| n_{A-1}, {\bf p}_n \rangle$ describes the 
 $(A-1)$-particle spectator system in the state $n$, with momentum ${\bf p}_n$. Note that, owing to NN correlations,  $| n_{A-1}, {\bf p}_n \rangle$ is not restricted to be  
a bound  state.

Within the IA, the contribution to the nuclear cross section arising from interactions involving the one-nucleon current 
is written in terms of the cross section of the elementary scattering process involving an individual nucleon and the nuclear spectral function $P({\bf k},E)$, 
dictating its energy and momentum distribution, according to   \cite{RMP}
\beq
\label{sigma1}
d\sigma_{IA} =  \ \int \,d^3k \ dE \ P(k,E)  \ d\sigma_{elem} \   . 
\eeq

\begin{figure}[h!]
\includegraphics[scale= 0.45]{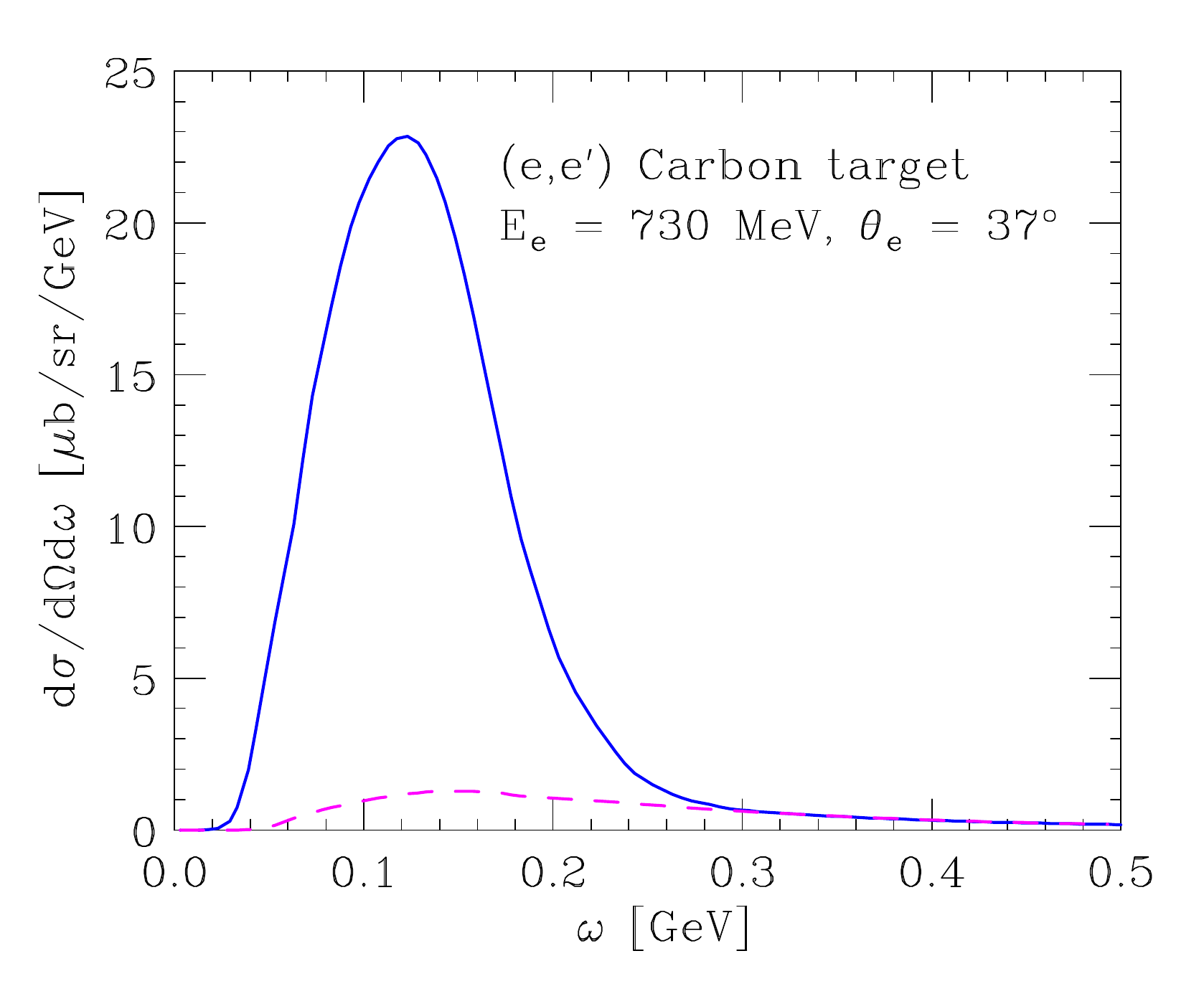} 
\vspace*{-.1in}
\caption{Cross section of the process $e+^{12}C \to e^\prime + X$ at beam energy $E_e= 730$ MeV and electron scattering angle $\theta_e= 37$  deg, 
plotted as a function of the energy loss. The solid and dashed lines represent the result of the full calculation and the contribution arising from 
amplitudes involving 2p2h final states.}
\label{ISC}
\end{figure}

Figure \ref{ISC} illustrates the 2p2h contribution to the electron-carbon cross section, at beam energy $E_e= 730$ MeV and  scattering angle $\theta_e= 37$  deg, arising 
from ISC. The solid line corresponds to the result of the full calculation, carried out within the IA using the spectral function of Ref. \cite{LDA}, while the dashed line has been 
obtained including only the amplitudes involving 2p2h final states.

The factorisation {\em ansatz} of Eq. \eqref{fact1} can be readily extended to allow for a consistent treatment of the matrix elements 
of one- and two-nucleon currents. The resulting expression is 
\beq
\label{fact2}
| N \rangle = | {\bf p} {\bf p}^\prime \rangle \otimes | m_{A-2} , {\bf p}_m \rangle \ ,
\eeq
where the states $| {\bf p} {\bf p}^\prime \rangle$ and $| m_{A-2} , {\bf p}_m \rangle$ describe two non interacting nucleons of momenta ${\bf p}$ and ${\bf p}^\prime$ and 
the $(A-2)$-particle spectator system, respectively.

Using Eq. \eqref{fact2}, the nuclear matrix element of the two-nucleon current can be written in terms of  two-body matrix elements according to 
\begin{align}
\label{matel:2}
\langle N | j_{ij}^{\mu} | 0 \rangle   =
 \int d^3k d^3 k^\prime   M_m({\bf k},{\bf k}^\prime&)  
\langle {\bf p} {\bf p}^\prime | j_{ij}^\mu | {\bf k} {\bf k}^\prime \rangle  \ ,
\end{align}
with the amplitude $M_m({\bf k},{\bf k}^\prime)$ given by
\begin{align}
\label{def:Mn}
M_m({\bf k},{\bf k}^\prime) = \left\{ \langle n_{(A-2)} , {\bf p}_m | \otimes \langle  {\bf k}  {\bf k}^\prime | \right\} | 0 \rangle \ .
\end{align}

Within the scheme outlined in Eqs. \eqref{fact2}-\eqref{def:Mn}, the nuclear amplitude $M_m({\bf k},{\bf k}^\prime)$ turns out to be independent of ${\bf q}$, 
and can therefore be obtained within non relativistic many-body theory without any problems. On the other hand, the two-nucleon matrix element can be 
evaluated using the fully relativistic expression of the current. 

The connection with the spectral function formalism 
becomes apparent noting that the two-nucleon spectral function $P({\bf k},{\bf k}^\prime,E)$,
yielding the probability of removing {\em two nucleons} from the nuclear ground state leaving the residual system with excitation energy $E$, is
defined as~\cite{spec2}
\begin{align}
\label{def:pke2}
P({\bf k},{\bf k}^\prime,E) = \sum_m |M_m({\bf k},{\bf k}^\prime)|^2 \delta(E + E_0 - E_m) \ ,
\end{align}
with $M_m({\bf k},{\bf k}^\prime)$  given by Eq. \eqref{def:Mn}.

\begin{figure}[h!]
\includegraphics[scale=0.45]{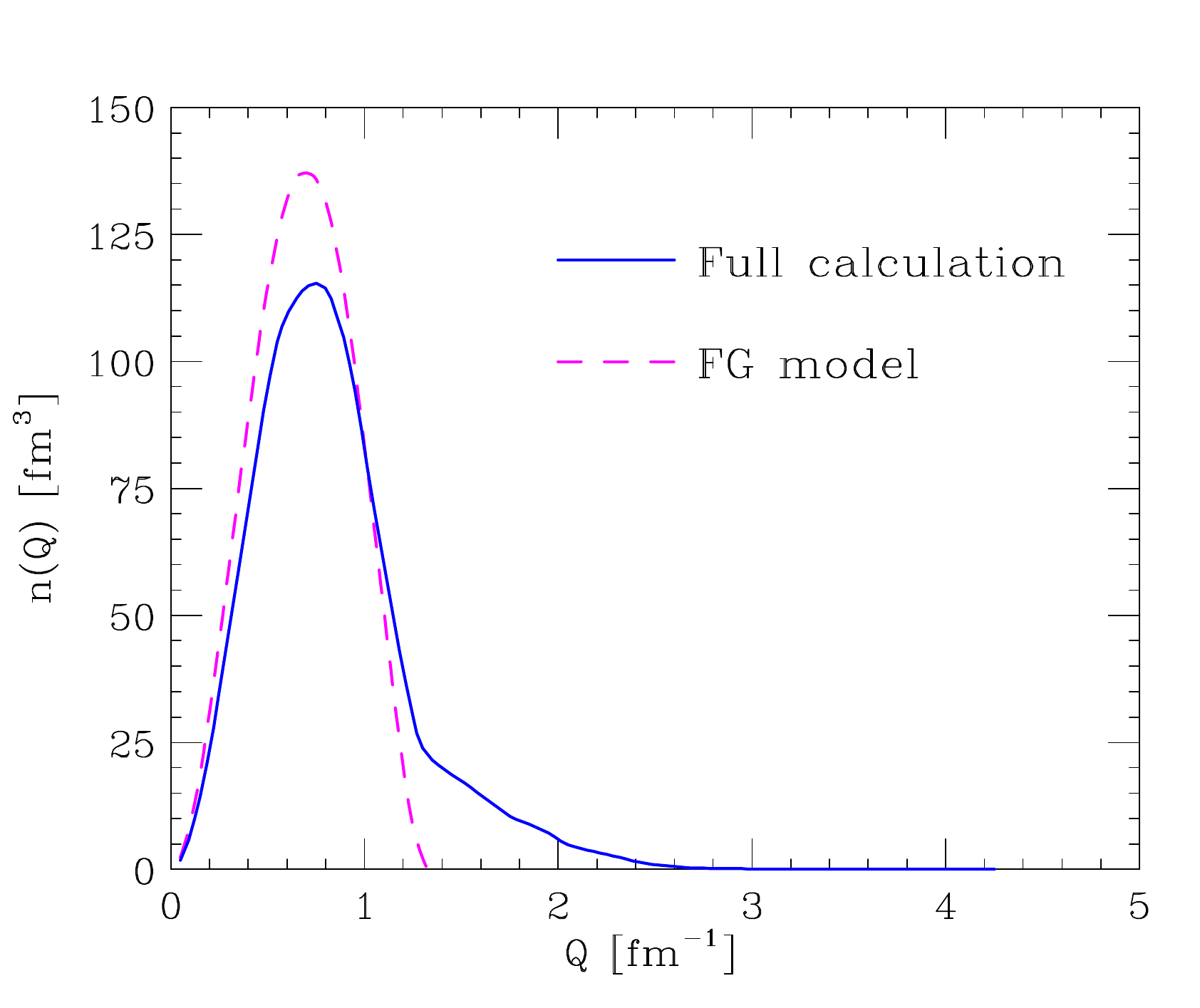}
\vspace*{-.1in}
\caption{Relative momentum distribution of a nucleon pair in isospin symmetric nuclear matter at equilibrium density.}
\label{SF}
\end{figure}

The two-nucleon spectral function of uniform and isospin symmetric nuclear matter at equilibrium density has been calculated 
 by the authors of Ref. \cite{spec2} using a realistic hamiltonian. The resulting relative momentum distribution, defined as
\begin{align}
\label{rel:dist}
n({\bf Q}) = 4 \pi |{\bf Q}|^2 \int d^3 K \ n\left( \frac{ {\bf Q} }{2} + {\bf K}, \frac{ {\bf Q} }{2} - {\bf K} \right)
\end{align}
where ${\bf K} = {\bf k}+{\bf k}^\prime$, ${\bf Q} = ({\bf k}-{\bf k}^\prime)/2$, and 
\begin{align}
n({\bf k},{\bf k}^\prime) = \int dE  \ P({\bf k},{\bf k}^\prime,E) \ ,
\end{align}
is shown by the solid line of Fig. \ref{SF}. Comparison with the prediction of the Fermi Gas (FG) model, represented by the dashed line,
 indicates that correlation effects are sizeable, and give rise to a quenching of the peak of the distribution, along with  the appearance
 of a high momentum tail.

Note that the {\em ansatz} of Eq. \eqref{fact1} implies neglecting {\em all} Final State Interactions (FSI) between the nucleon interacting with the beam 
particles and the spectators, including FSC.

In inclusive processes, FSI lead to: i)  a shift of the energy loss spectrum, arising from interactions between the knocked out nucleon and
the mean field of the recoiling nucleus, and ii) a redistribution of the strength from the quasi free bump to the tails, resulting
from FSC. Theoretical studies of electron-nucleus scattering suggest that in the kinematical region relevant to the MiniBooNE analysis
the former mechanism, which does not involve the appearance of 2p2h final states, provides the dominant contribution \cite{RMP}.
The inclusion of FSI within the IA scheme has been recently discussed  in Ref. \cite{benharFSI}.

The results presented in this Letter show that interference between the different reaction mechanisms leading to the excitation of 2p2h 
final states plays an important role, and must be taken into account using a description of nuclear structure that includes NN correlations.  
Models in which the processes involving MEC are treated within the framework of the IPM,  such as 
those of Refs.~\cite{martini,nieves}, appear to be conceptually inconsistent, although the impact of this issue on the numerical results 
needs to be carefully investigated. The treatment of MEC based on  the extension of the $y$-scaling analysis, extensively employed 
in electron scattering studies, also fails to account for interference effects \cite{barbaro}. 

The extension of the factorisation scheme underlying the IA appears to be a 
viable option for the development of a unified treatment of processes involving one- and two-nucleon currents in the region of large 
momentum transfer. We believe that the implementation of the approach outlined in this work may in fact be regarded as a first step 
towards the  {\em new paradigm} advocated by the authors of Ref. \cite{coletti}.

The authors are grateful to A. Ankowski, J. Carlson, S. Gandolfi, C. Mariani, S. Pieper and R. Schiavilla for many 
illuminating discussions. The work of OB and NR was supported by INFN under grant MB31. 
 NR gratefully acknowledges the hospitality of  the  Center for Neutrino Physics at 
 Virginia Tech, where part of this work was carried out.
Under an award of computer time provided by the INCITE program,
this research used resources of the Argonne Leadership Computing
Facility at Argonne National Laboratory, which is supported by the
Office of Science of the U.S. Department of Energy under contract
DE-AC02-06CH11357.

\end{document}